# ONTOLOGY-BASED EMERGENCY MANAGEMENT SYSTEM IN A SOCIAL CLOUD

Bhuvaneswari A[1] and Karpagam.G.R[2]

[1]Full Time Research Scholar, CSE Department, PSG College of Technology, Coimbatore, India
bhuvan888@yahoo.com

[2]Professor, CSE Department, PSG College of Technology, Coimbatore, India
grkrm@cse.psgtech.ac.in

### ABSTRACT

*The need for Emergency Management continually grows as the population and exposure to catastrophic failures increase. The ability to offer appropriate services at these emergency situations can be tackled through group communication mechanisms. The entities involved in the group communication include people, organizations, events, locations and essential services. Cloud computing is a "as a service" style of computing that enables on-demand network access to a shared pool of resources. So this work focuses on proposing a social cloud constituting group communication entities using an open source platform, Eucalyptus. The services are exposed as semantic web services, since the availability of machine-readable metadata (Ontology) will enable the access of these services more intelligently. The objective of this paper is to propose an Ontology-based Emergency Management System in a social cloud and demonstrate the same using emergency healthcare domain.*

## 1. INTRODUCTION

Emergency management is a procedure to manage the hazardous situations in a way to get out of the impact of disaster. Emergency Healthcare has gained importance due to the increase in awareness about health related problems. Nowadays Emergency Healthcare services [21] have been extended to pre-hospital care and transportation rather than being limited to the in-hospital treatment. Hence they expect efficient communication and rapid transportation. This expectation can be satisfied by group communication that involves entities like people, organizations, events, locations and essential services.

The need for consumer-driven adoption of new technologies and functionalities has led to the development of social networking applications such as Facebook, MySpace and Flickr, which have gained tremendous popularity [7]. These sites resemble a reporting tool of social events by enabling users to share their information like photos and videos worldwide. However they lack in social interaction required in Emergency Management like applications. The device and user interface constraints as well as slow data rates are the main barriers for interaction. Recent developments of i-phones and androids slash down such barriers enabling spontaneous interactions among people. The social interaction can be enabled in a social networking site by including web services. Still this may lead to resource constraints which can be resolved by Cloud Computing [17], a style of computing where massively scalable IT-related capabilities are provided "as a service" using Internet technologies. Such a concept in which Cloud Computing owned and maintained by a Social Network is called as Social Cloud. An economical and interactive Emergency Management System can be realized in a Social Cloud.

Another revolution in World Wide Web is Semantic Web which has gained the attention of lot of researchers. Semantic web describes methods and technologies to allow machine to understand the meaning or "semantics" of information on the World Wide Web using ontologies. Ontologies include computer-usable definitions of basic concepts in the domain and the relationships among them. They encode knowledge in a domain and also knowledge that spans domains. In this way, knowledge reusability is promoted by ontologies. The availability of machine-readable metadata would enable automated agents and other software to access the web more intelligently. The agents would be able to perform tasks and locate related information automatically on behalf of the user. Semantic search [13] locates information automatically by using the two types of ontologies namely domain ontology and service ontology. Domain Ontology is a conceptualization of a specific domain which enables the users to recognize the importance and relation between the terms and concepts of that domain. Service ontology is a conceptualization of set of services related to that particular domain. A third type of ontology, Cloud Ontology, a conceptualization of platforms, infrastructure and software that are supported and available in the cloud is introduced in this paper.

Ontology-based Social Cloud helps in explicating relationships among these entities. Ontological representation of Social Cloud has the significant advantages like relativity, validity and information gain.

1. Relativity: Social Cloud focuses on relationship among the platform, infrastructure, domain and services available in a cloud.

2. Validity: consistency of the ontology can be validated through reasoning and inference mechanisms; hence encoded information in a social cloud can be validated.

3. Information gain: ontologies and their inference mechanisms can come up with new relations and concepts from the existing ones among entities.

The provision of Emergency Healthcare system is a complex task which requires a set of services for completion. With the increasing popularity of service composition, these services can be used in a workflow to perform a task. Semantic Web Service Composition [1, 13] orchestrates existing services with the aid of ontologies which describe the domain of interest and web services involved. The workflow automation can be achieved by employing Fluent Calculus [2, 3, 8, 9], an Artificial Intelligence strategy. Fig. 1 depicts the domains that are involved in achieving the Ontology-based Social Cloud.

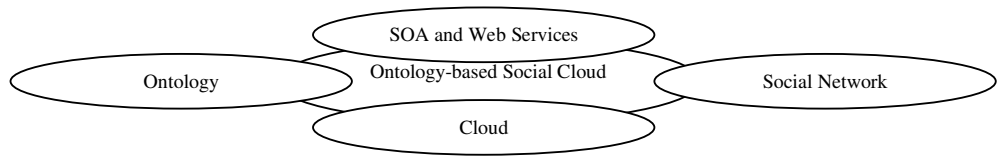

Fig. 1 Domains in Ontology-based Social Cloud

## 2. MOTIVATION

Travel has become an essential happening for every person and with comfort in mind mostly people plan journeys in train. A social network is a complex topology structure formed by the interactions between individuals in a society. Considering the thousands of passengers travelling everyday a social network can be set up among the passengers in a train. Thus Group communication that was utilized only for sharing the information among friends can be extended for efficient communication during emergency situations. For example the list of passengers will include both patients and doctors. With the establishment of a social network, emergency attention can be provided through rapid communication among people which will save the human life. This will create two major issues namely resource and time constraints. The resource constraints can be addressed by placing the social network in a cloud and time constraints by adding semantics to the cloud. A conceptual framework is proposed in this paper to realize the Ontology-based Emergency Management System.

## 3. CONCEPTUAL FRAMEWORK

The objective of the paper is to provide Ontology-based Emergency Management in a Social Cloud which provides an automated intelligent system to handle an emergency situation. To achieve this objective a layered architecture is shown in Fig. 2. The topmost Layer 1 explores the service, domain, infrastructure and platform ontologies that appropriately describe Ontology-based Social Cloud along with the business logic. Layer 2 emphasizes on AI strategies for workflow automation. Though there are several AI strategies such as Situation Calculus and Event Calculus, Fluent Calculus has been considered as a candidate as it resolves the frame problem and uses progression for planning. Layer 3 constitutes the application layer which consists of the application that accesses the above two layers. This paper considers the Emergency Healthcare System as the case study to demonstrate the Ontology-based Emergency Management System.

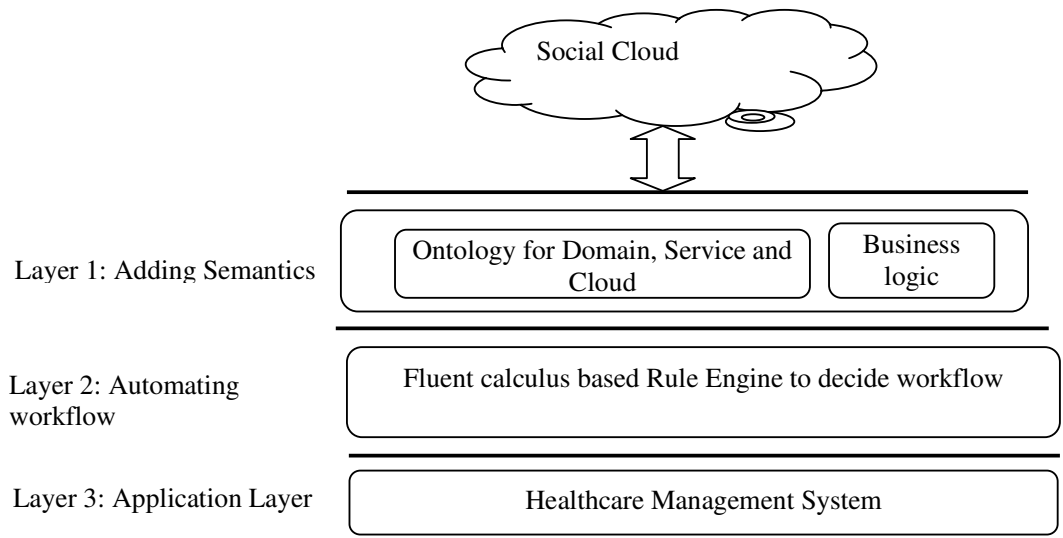

Fig. 2 Layered Architecture of Ontology- based Social Cloud

Ontology-based Social Cloud as shown in the Fig.2 is deployed in a private cloud which was set up using Eucalyptus [20], an open source software framework. The steps involved in setting the cloud are presented in Appendix.

### 3.1. Layer1- Adding Semantics

OWL Web Ontology Language-Description Logic (OWL-DL) was chosen as the ontology language as it has more facilities for expressing meaning and semantics. Protégé-OWL was utilized as the tool to build the proposed ontology. Protégé OWL[18] is a plug-in extension to the Protégé platform for building ontologies in OWL that aims to make Semantic Web technology available to a wide range of developers and users. OWL-S was employed to deploy a complete description of the semantic web services which supports the dynamic discovery, invocation, and composition of web services. The set of ontologies created for Ontology-based Social Cloud are cloud, domain and service ontology. The cloud ontology includes the details of the Infrastructure, Platform and Software that are provided by the cloud which is portrayed in Fig. 3

Fig. 3 Cloud Ontology

The scope of the domain ontology includes the generic concepts of Emergency Healthcare and their relationships that are applicable for health related risk occurred while the passenger is travelling in a train. It is intended to be used by emergency stakeholders like passenger (Patient/Medical Professional) and administrator for knowledge sharing and reuse. The required knowledge was acquired from the emergency domain experts, public reports, research journals and papers. A snapshot of the proposed domain ontology is depicted in Fig. 4.

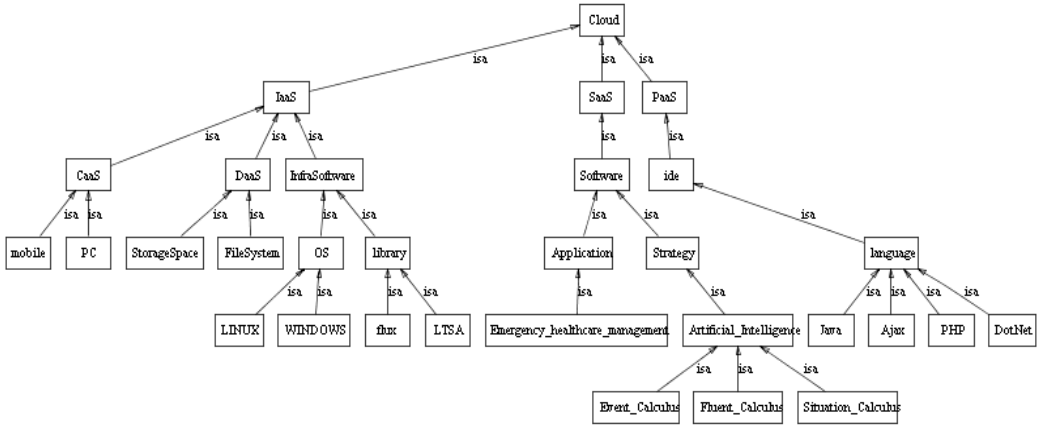

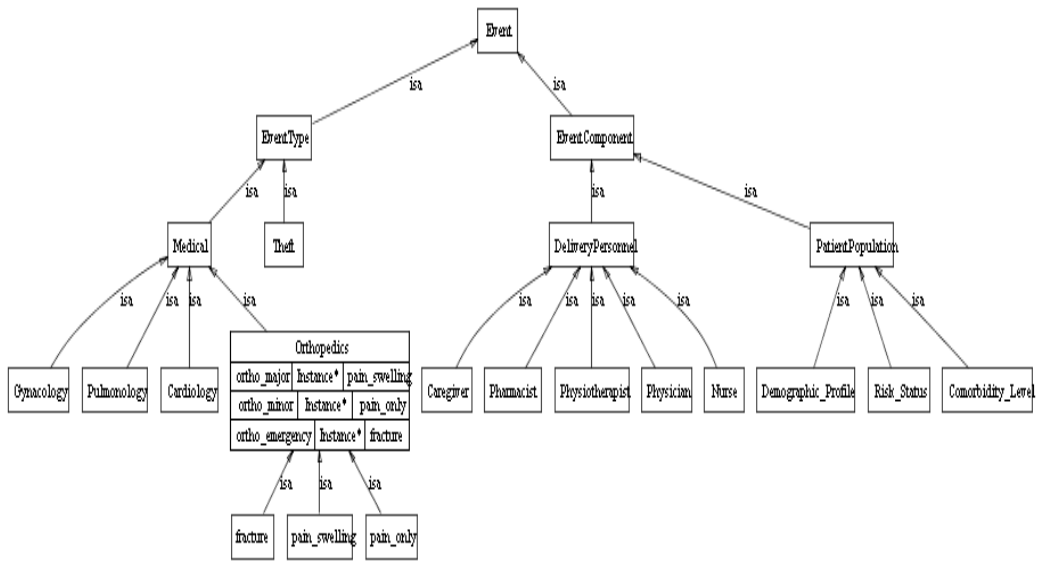

Fig. 4 Emergency Healthcare Domain Ontology

In this ontology, an event is an occurrence that has the potential to affect passengers either medical related or robbery related. Event has two main components that are EventType and EventComponent. EventType describes the type of event that had occurred either medical or robbery. EventComponent describes the participants of the event that includes patient population, delivery personnel and administrator. The EventType Medical is further classified according to the various specializations. Each specialization is classified according to the minor, major and emergency symptoms. In the proposed ontology orthopedics is classified as minor, if the patient has only pain; major, if the patient has pain with swelling and emergency if fracture.

OWL-S ontology [18] includes the profile, process and grounding information of the web services. The Service Profile provides a concise representation of web service capabilities (i.e. what the service does), through the advertising of the functionalities description; the Service Model gives a detailed description of how the service operates, specifically describing the transformations (i.e. the processes) that it undertakes; the Service Grounding supplies the details on how interoperate with a service, mapping the messages (according to the format and input/output specification provided in the process model) to the syntactic WSDL compliant form. The service ontology of a findResource service is shown in Table 1.

Table 1. Service Ontology Of findResource Service

| Service | <service:Service rdf:ID="FIND_RESOURCE_SERVICE"><br><service:presents rdf:resource="#FIND_RESOURCE_PROFILE"/><br><service:describedBy<br>rdf:resource="#FIND_RESOURCE_PROCESS_MODEL"/><br><service:supports rdf:resource="#FIND_RESOURCE_GROUNDING"/><br></service:Service> |
|---|---|
| Service Profile | <profile:Profile rdf:ID="FIND_RESOURCE_PROFILE"><br><service:isPresentedBy rdf:resource="#FIND_RESOURCE_SERVICE"/><br><profile:serviceName xml:lang="en"><br>FindResourceService<br></profile:serviceName><br><profile:textDescription xml:lang="en"><br>returns name and position of the resource for the required profession and<br>specialization<br></profile:textDescription><br><profile:hasInput  rdf:resource="#_PROFESSION"/><br><profile:hasInput rdf:resource="#_SPECIALIZATION"/><br><profile:hasOutput rdf:resource="#_NAME"/><br><profile:hasOutput rdf:resource="#_COACHNUM"/><br><profile:has_process rdf:resource="FIND_RESOURCE_PROCESS"<br>/></profile:Profile> |
| Process Model | <process:ProcessModel rdf:ID="FIND_RESOURCE_PROCESS_MODEL"><br><service:describes rdf:resource="#FIND_RESOURCE_SERVICE"/><br><process:hasProcess rdf:resource="#FIND_RESOURCE_PROCESS"/><br></process:ProcessModel><br><process:AtomicProcess rdf:ID="FIND_RESOURCE_PROCESS"><br><process:hasInput  rdf:resource="#_PROFESSION"/><br><process:hasInput  rdf:resource="#_SPECIALIZATION"/><br><process:hasOutput rdf:resource="#_NAME"/><br><process:hasOutput rdf:resource="#_COACHNUM"/><br></process:AtomicProcess><br><process:Input rdf:ID="_PROFESSION"><br><process:parameterType<br>rdf:resource="http://127.0.0.1/ontology/health.owl#Profession" /><br>  <rdfs:label></rdfs:label> |
| | `</process:Input>` <br> `<process:Input rdf:ID="_SPECIALIZATION">` <br> `<process:parameterType` <br> `rdf:resource="http://127.0.0.1/ontology/health.owl#Specialization" />` <br> `<rdfs:label></rdfs:label>` <br> `</process:Input>` <br> `<process:Output  rdf:ID="_NAME">` <br> `<process:parameterType` <br> `rdf:resource="http://127.0.0.1/ontology/health.owl#Name" />` <br> `<rdfs:label></rdfs:label>` <br> `</process:Output>` <br> `<process:Output  rdf:ID="_COACHNUM">` <br> `<process:parameterType` <br> `rdf:resource="http://127.0.0.1/ontology/health.owl#Coach" />` <br> `<rdfs:label></rdfs:label>` <br> `</process:Output>` |
| *Service Grounding* | `<grounding:WsdlGrounding rdf:ID="FIND_RESOURCE_GROUNDING">` <br> `<service:supportedBy rdf:resource="#FIND_RESOURCE_SERVICE"/>` <br> `</grounding:WsdlGrounding>` |

## 3.2. Layer 2 – Automating workflow

An AI planning problem [5] is generally described as a tuple **<$S$, $S_0$, $G$, $A$, $T$>** where: $S$ represents the set of all the possible states of the world; $S_0 \subset S$ is the initial word state; $G \subset S$ represents the goal state of the planner; $A$ represents the set of all possible actions; the translation relation $T \subset S$ x $A$ x $S$ defines the preconditions and effects for each action. Here $S$ x $A$ x $S$ is the set of all preconditions, an action and the set of all effects respectively. $T$ is a relation which is used to represent the transition of an action from one state to another based on precondition $\in$ S and effect $\in$ S. AI planning can be visualized as workflow automation of semantic web service composition. AI planning problem should comprise of descriptions for (i) initial state of the world, (ii) desired goal and (iii) set of actions to be executed in a formal language. This paper suggests Fluent Calculus as the formal language for workflow automation. Fluent Calculus is based on a set of axioms among which Action Precondition and State Update Axioms are essential to calculate the precondition and effects of the services. The Action Precondition axioms denote the precondition that should be satisfied to execute an action. The State Update Axioms define the relation between a state and its successor.

When health based emergency occurs the relevant delivery personnel should be identified and a notification must be sent. The precondition for finding the resource is the availability of a person's profession and specialization which implies that the required person is available. If the precondition is satisfied the person's name and coach number will be found by the findResource service. The precondition for notifying the resource is that identified person must be available. If the precondition is satisfied the notifyResource will send the message to that person. Table 2 shows the precondition and state update axioms for findResource and notifyResource services.

Table 2 Axioms in Fluent Calculus

| Service Name | findResource |
|---|---|
| Precondition Axiom | poss(findResource(PR,SP),Z):-<br>knows_val([PR],Profession(PR),Z),<br>knows_val([SP],Specialization(SP),Z),<br>holds(available(PR,SP),Z). |
| State Update Axiom | state_update(Z1, findResource(PR,SP),Z2) :-<br>holds(available(PR,SP),Z1),<br>update(Z1, [know(Name(P),know(CoachNum(CN)], [ ],Z2). |

| Service Name | notifyResource |
|---|---|
| Precondition Axiom | Poss(notifyResource(P,CN,MSG),Z):-<br>knows_val([P], Name(P),Z),<br>knows_val([CN],CoachNum(CN),Z),<br>knows_val([MSG], Message(MSG),Z),<br>holds(available(P,CN),Z). |
| State Update Axiom | state_update(Z1, notifyResource(P,CN,MSG), Z2):-<br>holds(available(P,CN),Z1),<br>update(Z1,[SendMsg(P,CN,MSG), know(ConfirmSend)],[ ],Z2). |

The verification of the Fluent Calculus can be done using LTSA (Labeled Transition System Analyzer) which is a verification tool for concurrent systems. It mechanically checks the specification of a concurrent system whether it satisfies the properties required for its behavior. The verification of the system and dynamism in Fluent Calculus had been clearly explained in [3]

### 3.3. Layer 3 – Application Layer

The details about the passenger like profession, disease, and other interests are collected while he is booking the ticket. A social network is created among these passengers using ontology. The passenger should validate his travel to avoid invalid registration. A contact number will be provided in the ticket to which the patient should make a call in an emergency. The admin who receives the call will report to the system about the emergency and information provided by the patient. The system will identify the resource to handle the emergency with the help of the ontology. The resource will acknowledge the request from admin, attend the patient and confirm the service completion to admin. Fig. 5 shows the actors and use cases identified for the above scenario.

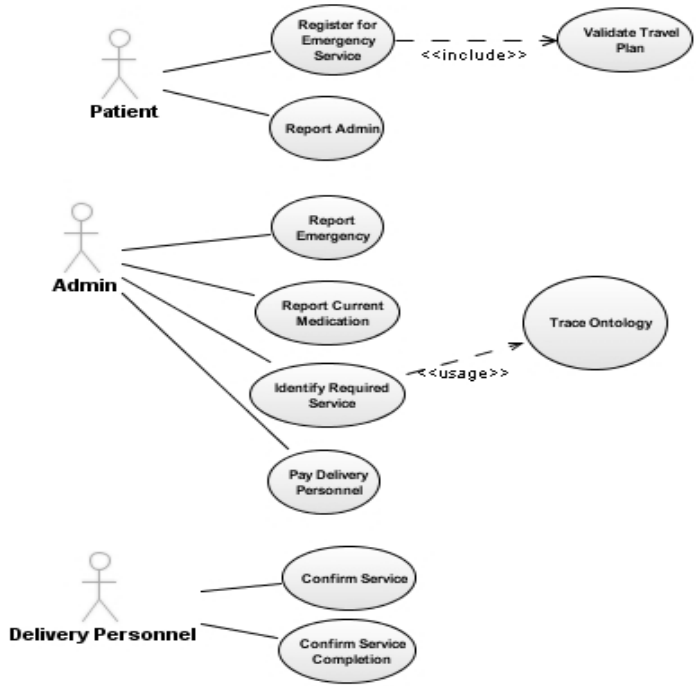

Fig. 5 Use Case Diagram – Emergency Healthcare System

Some of the essential use cases are provided with details like Name, Purpose, Description, Pre-condition, Post-condition, Basic flow and Alternate flow in the Tables 3, 4, 5, 6.

Table 3 Registration

| Name | Register for emergency service |
|---|---|
| Purpose | Passenger registers his personal and medical details to avail or provide (if Delivery Personnel) emergency service |
| Description | When the passenger reserves a ticket for his journey, an option is provided for him to utilize or provide emergency service. If he is interested in utilizing that service he has to fill up the form which requires specific illness, medication undergone and medicine-in-hand. In case of delivery personnel he must register his profession. |
| Precondition | The passenger must have reserved the ticket |
| Post condition | Medical details of the registered passenger will be updated in social network ontology |
| Basic flow | 1. Reserve the ticket<br>2. Select the option for registration<br>3. If option = yes then<br>   registration form is displayed;<br>4. The registration form is filled up.<br>5. Ticket is confirmed and ontology is updated. |
| Alternate flow | 1. If option = no then<br>   ticket is confirmed;<br>   "no medical  Service" is prompted;<br>2. Quit the application |

Table 4 Validate Travel Plan

| Name | Validate Travel Plan |
|------|----------------------|
| Purpose | The Passenger who registers for emergency service should validate his travel plan |
| Description | The registered user of emergency service should confirm his travel by validating the departure and arrival places, date of journey. |
| Precondition | The passenger must have reserved the ticket |
| Post condition | Journey of the passenger is validated |
| Basic flow | 1.    Ticket is reserved for travel.<br>2.    The registration form is filled up.<br>3.    Travel plan will be updated.<br>4.    Ticket is confirmed and ontology is updated. |

Table 5 Report Emergency

| Name | Report Emergency |
|------|------------------|
| Purpose | The Admin should intimate the system about the occurrence of emergency |
| Description | When the registered user informs about the need for emergency service admin will intimate the system about Emergency. |
| Precondition | The passenger must have registered for emergency service; |
| Post condition | Care provided to the patient ;<br>Date, Time, Patient Name, Case History, Coach No., Seat No and Delivery Personnel are stored. |
| Basic flow | 1. Admin receives information about Emergency<br>2. Admin checks whether the passenger had registered for Emergency Service<br>3. If registered = Yes then<br>   Admin alerts the system<br>4. Delivery Personnel is alerted by the system |
| Alternate flow | 1.   If registered = No then<br>      Permit the patient to register<br>2.   Payment collected by the Admin |

Table 6 Trace Ontology

| Name | Trace Ontology |
|------|----------------|
| Purpose | To categorize the list of medical related persons from the network |
| Description | Categorizes the medical related persons according to their profession first and then according to their relative position from the patient |
| Precondition | The persons must have been listed according to the major category, medical, for example |
| Post condition | A list is generated with all possible categorization |

| Basic flow | 1. A list of persons identified is given as input |
| | 2. Categorization done based on profession |
| | 3. If profession is doctor categorize according to specialization |
| | 4. Categorization done based on coach number |
| | 5. Display the sorted list to the admin |
| Alternate flow | If no persons available under major category then the intimation must be sent to next nearest Railway Station's authority |

# 4. RELATED WORK

The service composition is defined as the process of combining and linking existing web services to assemble new web processes, when no atomic web service can fulfill the user's requirements. [7] discusses about the automated web service composition methods and concludes that workflow based automation is useful when the process model is provided and the AI planning based automation is useful when the requester has no process model but has a set of constraints and preferences. Composing the syntactically described web services is a difficult task since only inputs and outputs are specified in Web Service Description Language (WSDL). Semantically described web services will have the precondition and effects of the service in addition to inputs and outputs. Hence semantic web service composition has gained a lot of attention by various researchers [4, 5, 14].

Reasoning about actions to be performed is a major requirement of any autonomous agent trying to achieve goals. This can be achieved by first order logic categorized under Logic Programming of Artificial Intelligence. First order logic has a subset of calculi namely Situation Calculus, Event Calculus and Fluent Calculus. Situation Calculus suffers by the frame problem. This can be solved by introducing an additional state update axiom in fluent calculus and by using circumscription in Event Calculus [11]. Fluent Calculus [9, 10] can be experimented by using the language FLUX which is based on constraint logic programming and addresses the frame problem whereas Event calculus is not having any specific language. Hence it can be concluded that fluent calculus is the best candidate for AI-based planning problems which has been explored in detail in the previous papers [2,3]. In [15] authors have shown that combination of RDF models and ontologies provide the best means to model and visualize social networks. They had worked only with small and well defined social networks and had not mentioned about the extensibility of the network. [6] introduced a new network application scenario for social interaction – the mobile ad hoc social network, MobiSN. The MobiSN system will eventually revolutionize the way humans may interact with one another, by removing the barriers resulting through unfamiliarity. [12] had proposed an ontology based knowledge base developed for tracking terrorism related data on the web.

A lot of research work is being carried out regarding the utilization of social network to serve the public. This paper involves deploying a private cloud and experiments the interoperability of semantic web services needed for service composition in a cloud setup.

# 5. CONTRIBUTION

In order to investigate the use of semantic web service composition in a social cloud, the following works have been carried out.

1. Web services are integrated with the private cloud hosted by Eucalyptus with the server placed on the cloud controller. The image started should have all the basic packages needed by application such as apache tomcat, java-sdk, etc. The server will get request from the user who will be viewing the jsp page hosted by apache tomcat running on the cloud. It will run a query match against ontology and return results which are sent back to the user.

2. The ontologies are kept and maintained at server side on cloud and can be updated using an AI trigger, whenever a need arises like alterations or updations of the offered services. Ontology provides metadata about the cloud, domain of interest and web services which enables the cloud interoperability in a new dimension.

3. The group communication required in an emergency situation is enhanced by setting up a social cloud.

This paper has concentrated only on the automation of service composition in a private cloud setup and dynamism has been evidenced from [3]. The dynamism in Healthcare domain should be exploited with the same scenario in future.

## 6. DISCUSSION AND CONCLUSION

The experimental study reveals that integrating cloud, semantic web services and social network will enhance the pervasive group communication. A private cloud setup has been deployed and the services are executed in the cloud. Cloud interoperability is provided at metadata level by ontology as it provides the complete description about the cloud, emergency healthcare domain and set of services involved. In the case study, support for automation has been evidenced through Fluent Calculus based workflow generation. In future the same work will be extended to handle dynamism and experience the scenario in a public cloud environment.

# APPENDIX – PRIVATE CLOUD SET-UP PROCEDURE

The Ubuntu Enterprise Cloud (UEC) setup includes two servers (Server 1 and Server 2) which will run a Lucid 64-bit server version and the third system which will run a Lucid Desktop 64-bit version (Client)

**Installation Process for Server 1**

• Boot the server off  the Ubuntu Server 10.04 CD. At the graphical boot menu, select "Install Ubuntu Enterprise Cloud" and proceed with the basic installation steps.
• Installation only lets you set up the IP address details    for   one interface. Do that for eth0.
• We need to choose certain configuration options for UEC, during the course of the install.
• Cloud Controller Address - Leave this blank as Server1 is the   Cloud Controller in this setup.
• Cloud Installation Mode - Select "Cloud controller", "Walrus storage service", "Cluster controller" and "Storage controller".
• Network interface for communication with nodes - eth1
• Eucalyptus cluster name – cluster1
• Eucalyptus IP range - 192.168.4.155-192.168.4.165

**Installation Procedure for Server 2**

Repeat the same procedure as in server 1 with
eth0 configured as 192.168.4.146

- Cloud Controller Address - 192.168.4.145
- Cloud Installation Mode - Select "Node Controller"
- Gateway - 192.168.4.145 (IP of the CC)

**Installation Procedure for Client**

The purpose of Client machine is to interact with the cloud and can also be used for bundling and registering new Eucalyptus Machine Images (EMI).
- Boot the Desktop off the Ubuntu Desktop 10.04 CD and install.
- Install KVM to help us install images on KVM platform and bundle them:

$apt_get install qemu_kvm

**Accessing the cloud**

- Login to the cloud controller by using the following link https://192.168.4.145:8443. The default username is "admin" and the default password is "admin".
- Note that the installation of UEC installs a self signed certificate for the web server. The browser will warn us about the certificate not having been signed by a trusted certifying authority. Authorize the browser to access the server with the self signed certificate.
- When you login for the first time, the web interface prompts to change the password and provide the email ID of the admin. After completing this mandatory step, download the credentials archive from https://192.168.4.145:8443/ #credentials and save it in the ~/.euca directory.

**Checking Availability of Resources**

- Extract the credentials archive:

$ cd .euca
$ unzip mycreds.zip
- Source eucarc script to make sure that the environmental variables used by euca2ools are set properly.
   $ . ~/.euca/eucarc
- To verify that euca2ools are able to communicate with the UEC, try fetching the local cluster availability details shown below in Fig. 6.

$ euca-describe-availability-zones verbose

Fig. 6 List of Available Resources

## RUNNING INSTANCES

**Installing Cloud Images**

No image exists by default in the Store (web Interface). Running an instance or VM in the cloud is only based on image. Images can be installed directly from Canonical online cloud image store or we can also build custom image, bundle it, upload and register them with the cloud. The "Store" tab in the web interface will show the list of images that are available from Canonical over the internet as shown in Fig. 7.

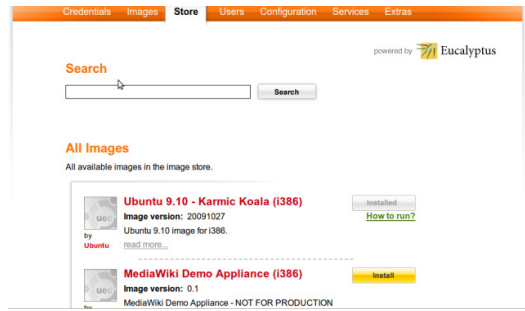

Fig. 7 List of Images from Store

## a. Checking Images

"euca-describe-images" is the command-line equivalent of clicking the "Images" tab in the Eucalyptus administrative web interface. This shows the emi-xxxxxx identifier for each image that are present in cloud.

$ euca-describe-images

## b. Creating a Keypair

Build a keypair that will be injected into the instance allowing us to access it via ssh:

$ euca-add-keypair mykey > ~/.euca/mykey.priv

$ chmod 0600 ~/.euca/mykey.priv

## c. Running the Instances

start an instance of the image and it can be accessed through ssh

$ euca-run-instances -g Ubuntu 9.10 -k mykey -t c1.medium emi-E088107E

After issuing the "euca-run-instances" command to run an instance, we can track its state by using the euca-describe-instances command and the output is shown below in Fig. 8.

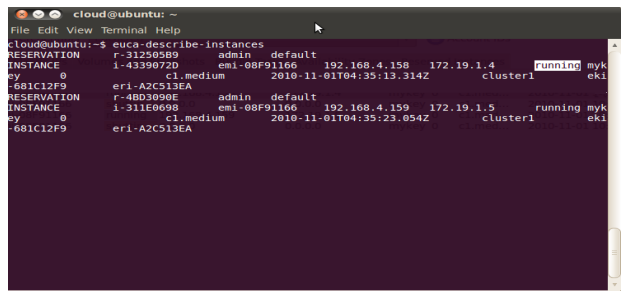

Fig. 8 Snapshot of Running Instances

To integrate the reasoner with the private cloud hosted by Eucalyptus, the server will be on the cloud controller. The image started should have all the basic packages needed by application such as apache tomcat, java-sdk etc. The server will get request from the user who will be viewing the jsp page hosted by apache tomcat running on the cloud. It will run a query match against ontology and return results which are sent back to the user. The ontologies are kept and maintained at server side on cloud and can be periodically updated depending on the new developments and additions in services offered.

## Authors

Ms.A.Bhuvaneswari is a Full time Research candidate pursuing her Research in Department of Computer science and Engineering in PSG College of Technology affiliated to Anna University. Her area of interest includes Bio Inspired Computation and Web Service Composition. Her research focuses on investigating the AI techniques for planning phase of Semantic Web service Composition. She has published papers in IEEE and ACM digital libraries and National/International Journals.

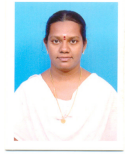

Dr.G.R.Karpagam is Professor with 15 years of experience in Department of Computer science and Engineering in PSG College of Technology. She is an IEEE member interested in the field of web services, Model Driven Architecture and cloud service discovery. She has published papers in National /International Journals and Conferences and reviewer of International Journals She has co-authored monograph for Database management systems and has assisted in writing ISO Procedure manual. She has established Industry funded laboratories namely Open-Source Software Laboratory, Service Oriented Architecture Laboratory and Cloud Computing Laboratory in Collaboration with leading MNCs.

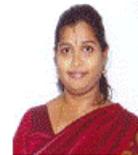